\documentclass[prl,superscriptaddress,letter,10pt,twocolumn]{revtex4}
\usepackage{amsmath}
\usepackage{amssymb}
\usepackage{graphicx}
\usepackage{color}
\usepackage{epsfig}
\usepackage{dcolumn}
\usepackage{bm}
\usepackage{bm}
\usepackage{blindtext}
\usepackage{natbib}
\usepackage{epstopdf}
\usepackage{ulem}
\usepackage{url}
\usepackage{bm}
\usepackage{subfigure}

\def\be{\begin{equation}}
\def\ee{\end{equation}}
\def\bea{\begin{eqnarray}}
\def\eea{\end{eqnarray}}

\def\ba{\begin{aligned}}
\def\ea{\end{aligned}}

\newcommand{\T}{\mathbb{T}}
\newcommand{\Z}{\mathbb{Z}}

\begin{document}
\title{Stability and fine structure of symmetry-enriched quantum criticality
in a spin ladder triangular model}
\author{Xiao Wang}
\thanks{These authors contributed equally to the work.}
\affiliation{Tsung-Dao Lee Institute, Shanghai, 201210, China}
\author{Linhao Li}
\thanks{These authors contributed equally to the work.}
\affiliation{Institute for Solid State Physics, The University of Tokyo. Kashiwa, Chiba 277-8581, Japan}
\author{Jianda Wu \footnote{Electronic address: wujd@sjtu.edu.cn}}
\affiliation{Tsung-Dao Lee Institute, Shanghai, 201210, China}
\affiliation{School of Physics \& Astronomy, Shanghai Jiao Tong University, Shanghai, 200240, China}
\affiliation{Shanghai Branch, Hefei National Laboratory, Shanghai 201315, China}

\begin{abstract}
In this letter, we propose and study a ladder triangular cluster model which
possesses a $\mathbb{Z}_2$ symmetry and an anti-unitary $\mathbb{Z}^{\T}_2$ symmetry
generated by the spin-flip and complex conjugation, respectively. The phase diagram of
the model hosts a critical line between a spontaneous symmetry breaking phase and
a symmetry protected topological phase.
Along the critical line, one endpoint exhibits symmetry-enriched Ashkin-Teller universality (SEATU),
while other critical points fall into the symmetry-enriched  Ising universality (SEIU). Both universality classes accommodate symmetry protected degenerate edge modes under open boundary conditions.
This degeneracy can be lifted with a gap opening when proper perturbation is applied to the boundary.
With system size ($L$) increasing, at the point of SEATU,
the gap closes following $L^{-1}$. In contrast, for the critical points of SEIU
apart from a point with the known gap
closing as $L^{-14}$, other points surprisingly show
exponentially gap closing.
The coexistence of different gap closing behaviors
for critical points of
the same symmetry-enriched universality goes beyond the
the usual understanding of symmetry-enriched universality class, implying a fine and rich structure of phase transition and universality class.

\end{abstract}
\maketitle

\textit{Introduction.---}Symmetry and topology play fundamental roles
in fascinating emergent phenomena of quantum many body physics. Notably, the short-range entangled gapped phases
can be classified by incorporating specific symmetries, known as symmetry
protected topological (SPT) phases \cite{PhysRevB.82.155138}. In the phase diagram, two distinct SPT phases cannot be connected by a smooth path that maintains symmetry and does not encounter phase transitions.
The most general classification of ($d$+1)D SPT phases protected
by an on-site bosonic symmetry group $G$ is based on the group cohomology $H^{d+1}(G,U(1))$ \cite{PhysRevB.86.115109,PhysRevLett.114.031601,PhysRevB.87.155114,chen2012symmetry,PhysRevB.90.235137,PhysRevB.89.035147}.
In particular, when $d$=1, the nontrivial SPT order is manifested on the existence of exponentially
localized zero-energy edge modes under open boundary conditions (OBC),
which carry nontrivial projective representations of $G$ \cite{PhysRevB.84.165139,PhysRevLett.59.799,PhysRevB.83.035107,PhysRevB.81.064439,PhysRevB.85.075125}.
Such edge modes cannot be removed when the bulk is gapped and symmetry-preserved.

Recently, more attentions are paid to systems exhibiting quantum critical points (QCPs)
with continuous quantum phase transitions (QPTs).
The QCP with continuous QPT is essentially the renormalization group fixed points,
which enjoys scaling invariant and can be
categorized by the concept of universality class~\cite{sachdev_2011}.
It was further realized that quantum critical states of the
same universality class can be classified into distinct subclasses
with certain symmetries imposed, if they can not be connected by a smooth
path without going through a multi-critical point~\cite{verresen2021gapless,longzhang_2022,fn_smoothpath}. This yields the notion of symmetry-enriched quantum criticality (SEQC), 
which is the analogue of SPT phases in critical systems~\cite{PhysRevX.7.041048,PhysRevX.11.041059,PhysRevB.97.165114,PhysRevB.104.075132,PhysRevLett.122.240605,yang2022duality,li2022symmetry,Li:2023mmw,PhysRevB.106.144436}.
Some SEQC can be distinguished by the non-local symmetry flux operators or the string order parameters, which serves as a topological invariant and can imply robust degenerate edge states.

\begin{figure}
	\includegraphics[width=7.5cm]{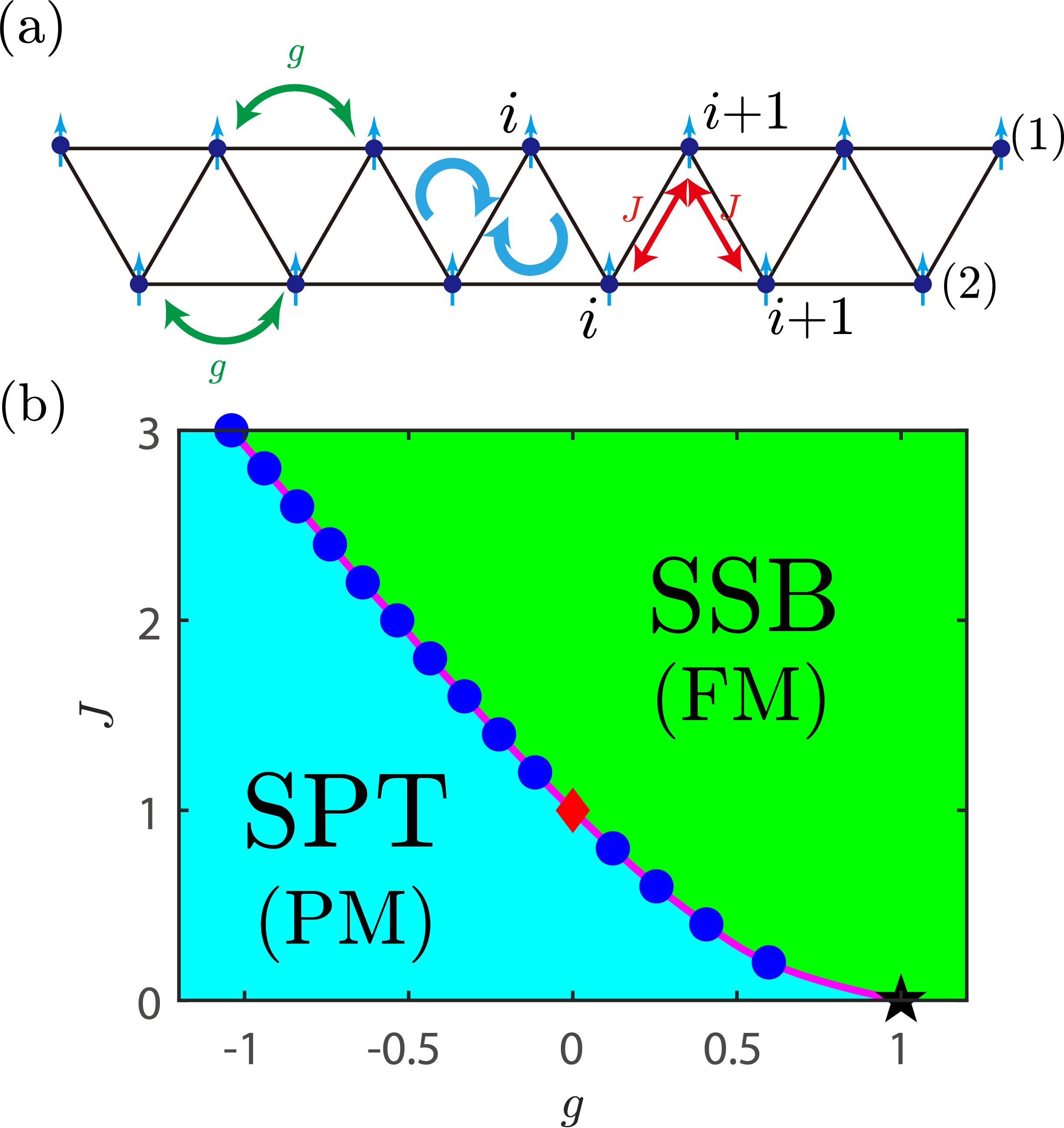}
	\caption{(a) An LTC model with intra-chain coupling $g$, inter-chain coupling $J$, and cluster coupling (labeled as cyan color). (b) The phase diagram of the LTC model. $(g,J)=(1,0)$ in an AT universality class is labeled as a black star. The middle point in the symmetry-enriched Ising universality class $(g,J)=(0,1)$ which shows an algebraic energy gap closing behaviour is labeled as a diamond.}
	\label{fig_model}
\end{figure}

While the SEQC in one-dimensional (1D) quantum many-body systems
has attracted a plethora of studies,
in this letter
we take one step further investigation
by focusing on studying a ladder triangular cluster (LTC) model Eq.~(\ref{Eq:cluster}) as
illustrated in Fig.~\ref{fig_model}(a)].
The model hosts a nontrivial SPT phase and a $\mathbb{Z}_2$ spontaneous symmetry breaking (SSB) phase~[Fig.~\ref{fig_model}(b)]. By tuning the inter-chain and intra-chain coupling,
a critical line between SPT and SSB phases is determined.
Using infinite time evolving block decimation (iTEBD) algorithm~\cite{vidal_tebd},
it is found that one endpoint [the black star in Fig.~\ref{fig_model}(b)] of the critical curve is described by symmetry-enriched
Ashkin-Teller university (SEATU) class~\cite{OSHIKAWA1997533,PhysRevLett.77.2604,ginsparg1988applied},
while all other critical points belong to the symmetry-enriched Ising universality (SEIU) class~\cite{verresen2021gapless,Ising_review}.
Both the two classes display SEQC, and
possess a string order parameter with nonzero
scaling dimension, and accommodate an exact four-fold (SEATU) or
two-fold (SEIU) degenerate ground state with OBC.
Furthermore, proper symmetric boundary perturbations
to SEQC models with OBC may lift the degeneracy, resulting in an energy gap.
The gap exhibits either exponential or algebraic (power law)
closing behavior with system size increasing~\cite{verresen2021gapless}.
In the case of the QCP of SEATU the gap closing appears to follow $1/L$.
However, for the QCPs of SEIU,  it is surprising to find that, other than a point [the red diamond in Fig.~\ref{fig_model}(b)]
exhibiting known $1/L^{14}$ gap closing behavior \cite{PhysRevX.11.041059},
all other points display exponentially gap closing behaviour.
The coexistence of different gap closing behaviors for the QCPs
within the same symmetry-enriched universality class reveals
fine structure lurking in the symmetry-enriched universality, which
goes beyond conventional understanding in SEQC.

\textit{Model and Phase Diagram.---}The Hamiltonian of LTC model [Fig.~\ref{fig_model} (a)] follows
\be
H_{\text{LTC}} = H_{\text{C}} + H_{\parallel} + H_{\perp},
\label{Eq:cluster}
\ee
with
\be
H_{\text{C}} = -\sum^{N}_{i=1}Z^{(1)}_{i-1} X^{(2)}_{i-1} Z^{(1)}_{i} + Z^{(2)}_{i} X^{(1)}_{i+1} Z^{(2)}_{i+1},
\label{Eq:clustercoupling}
\ee
\be
H_{\parallel} = -g\sum^N_{i=1} \left(Z^{(1)}_i Z^{(1)}_{i+1} + Z^{(2)}_i Z^{(2)}_{i+1}\right),
\label{Eq:intrachain}
\ee
and
\be
H_{\perp} = -J\sum^N_{i=1}\left( Z^{(1)}_{i} Z^{(2)}_{i}+ Z^{(2)}_{i}Z^{(1)}_{i+1}\right),
\label{Eq:interchain}
\ee
where $A^{(m)}_{i},A=X,Y,Z$ are Pauli matrices on site $i$ in the chain $m=1,2$,
and $g$ and $J$ are intra-chain and inter-chain couplings, respectively.
The cluster Hamiltonian $H_{\text{C}}$ carries the main contribution of nontrivial SPT phase.
$H_{\text{LTC}}$ respects an anti-unitary symmetry following complex conjugation $K$,
and a $\mathbb{Z}_2$ symmetry under spin-flip operation $U=\prod_i X^{(1)}_i X^{(2)}_i$.
When $g, J \ll 1$, the model is dominated by $H_C$,
staying in the nontrivial SPT phase with non-zero string order parameter \cite{Chen2013SymmetryprotectedTP}.
When $g, J \gg 1$, the strong Ising interaction drives the model
into the SSB phase.
Furthermore, by tuning the coupling parameters $g$ and $J$, the model hosts
a phase transition between a nontrivial SPT \cite{PhysRevLett.50.1153,PhysRevLett.59.799,PhysRevB.40.4709}
and an SSB phases [Fig.~\ref{fig_model} (b)],
corresponding to paramagnetic (PM) and ferromagnetic (FM) phases, respectively.

By applying the decorated domain wall (DW) transformation,
\be
U_{\text{DW}}=\prod^{N}_{i=1}\exp\left(\frac{\pi i}{4}(2-Z^{(1)}_{i}Z^{(2)}_{i}+Z^{(2)}_{i}Z^{(1)}_{i+1})\right).
\label{Eq:UDW}
\ee
The LTC model can be mapped to a simpler triangular transverse field Ising ladder (TTFIL),
\be
\ba
H_{\text{TTFIL}} = -\sum^{N}_{i=1}\left( X^{(1)}_{i} + X^{(2)}_{i+1} \right) + H_{\parallel}+H_{\perp},
\label{Eq:TIL}
\ea
\ee
where the gapped SPT phase of the LTC model is related to a trivial gapped phase of
the TTFIL model by the $U_{\text{DW}}$ transformation~\cite{Chen2013SymmetryprotectedTP,PhysRevB.91.155150,PhysRevB.106.224420}.
The TTFIL model shares the same critical line with the LTC model,
which is more convenient for
determining the critical line and universality class.

Now we go into details of the TTFIL model and further investigate
the universality class for the two models along the critical line correspondingly.
When $J=0$, the TTFIL model accommodates $\mathbb{Z}^{(1)}_2 \times \mathbb{Z}^{(2)}_2$ symmetry
which is generated by spin flip on each independent chain \cite{ladderTFIC}. The Hamiltonian reads
\be\label{eq:TIL Ham}
H_{\text{TTFIL}}=-\sum_{m=1,2}\sum_{i=1}^{N}\left( X_{i}^{(m)}+g Z_{i}^{(m)}Z_{i+1}^{(m)}\right),
\ee
which contains two independent transverse field Ising chains. For each chain,
the QCP between the trivial gapped phase and the SSB phase appears at $g=1$,
corresponding to the Ising universality class.
Thus the TTFIL at $g=1$ is in an Ashkin-Teller university (ATU) class
with central charge $c=1$
\cite{IsingCFT1989,LECLAIR1998523,BANKS1976119,ladderTFIC}.
Correspondingly, the LTC model is in a SEATU class
at the critical point $(g, J)= (1, 0)$.

When tuning on the inter-chain coupling $J$, the
$\mathbb{Z}^{(1)}_2 \times \mathbb{Z}^{(2)}_2$ symmetry breaks
into the diagonal $\mathbb{Z}_2$ symmetry~\cite{ladderTFIC}
(generated by $U$ as aforementioned).
By using iTEBD algorithm, it is further found that all the other critical points with $J>0$ fall into Ising universality class, implying the corresponding critical points of LTC model belongs to symmetry-enriched Ising universality class~\cite{ladderTFIC}. Importantly, since the central charge remains invariant along the critical line as long as $J>0$, the system should be in the same SEQC, based on the definition provided by the concept of a smooth path. However, we will show there are two distinct subclasses in SEIU depending on the boundary properties, which is beyond the conventional understanding of SEQC above.

\begin{figure*}
	\includegraphics[width=17cm]{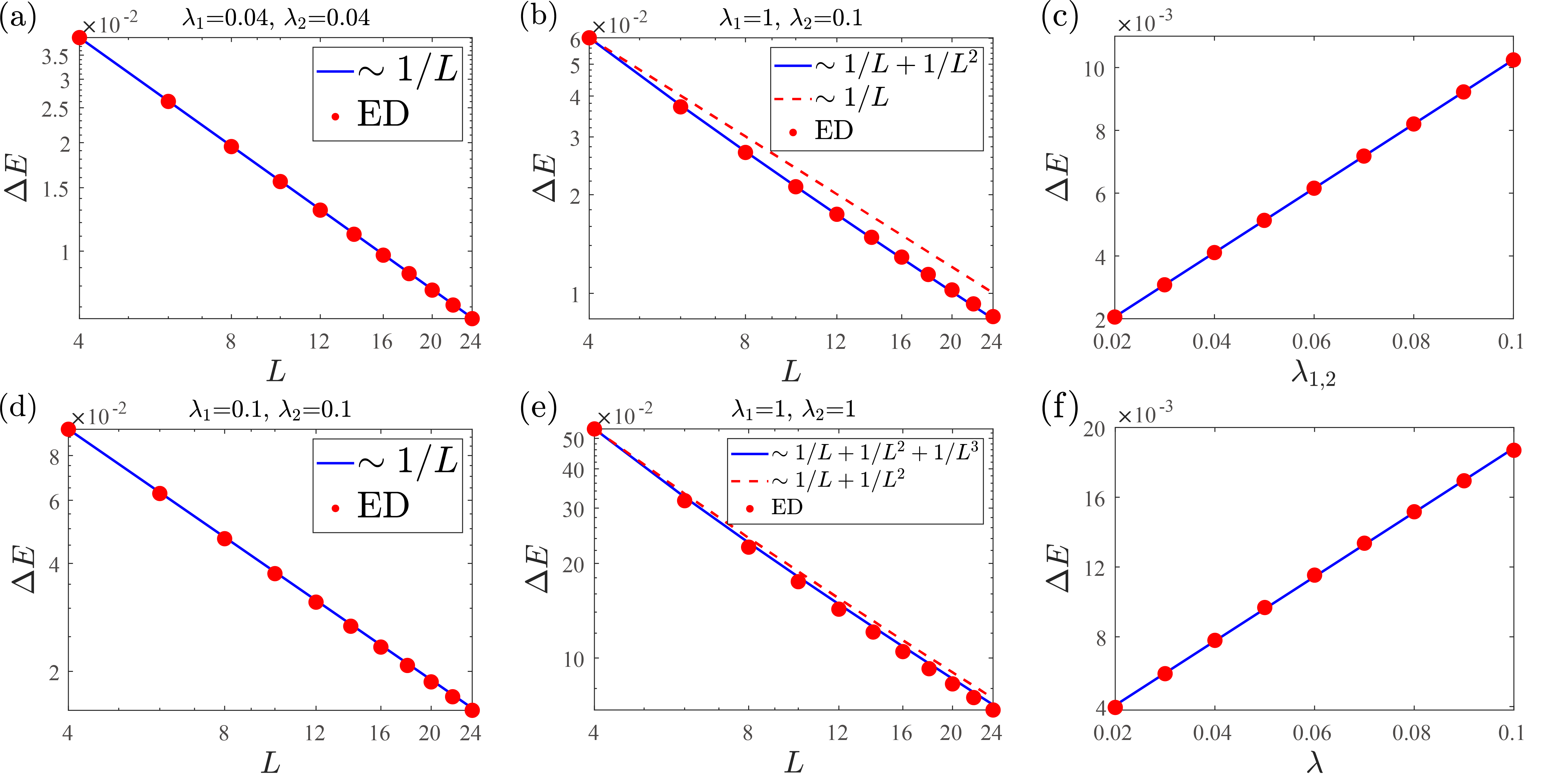}
	\caption{(a) (b) (d) and (e) Energy splitting $\Delta E$ as a function of system size $L=2N$. (a) and (d) $\Delta E(L)$ with chosen parameter $\lambda_{1,2}=0.04,0.1$, the $1/L$ closing behaviour greatly fits the data from exact diagonalization. (b) and (e) $\Delta E(L)$ with chosen parameter $\lambda_{1}=1,\lambda_{2}=0.1$ and $\lambda_{1}=1,\lambda_{2}=1$ respectively. It is shown that with increasing $\lambda$, higher order corrections need to be further considered. (c) $\Delta E(\lambda_{1}=\lambda_{2})$ and (f) $\Delta E(\lambda_{1},\lambda_{2}=1)$ with fixed $L=20$. The energy splitting also follows a linear behaviour as a function of $\lambda_{1,2}$.}
	\label{fig2}
\end{figure*}

\textit{Stability of the SEQC.---} For (1+1)D nontrivial SEQC, 
it is known that there are multiple (quasi)-degenerate ground states with 
finite size gap  under OBC. In the case of the LTC model, this property could be 
understood through the two-fold exact ground state degeneracy 
at finite size with OBC, corresponding
to spontaneous boundary magnetization \cite{verresen2021gapless}. 
This comes from the anti-commutation relation between the spin operators on the boundary $\{Z^{(1)}_{1},Z^{(2)}_{L}\}$ and $\Z_2$ symmetry operator,
while both of them commute with the Hamiltonian.

Moreover, symmetric boundary perturbations can be applied to detect the stability 
of such ground-state degeneracy. In general, once adding the perturbation, 
the exact degeneracy is broken and the finite size energy splitting is triggered. 
In the thermodynamic limit, such energy splitting shows exotic exponentially or algebraically 
(power law) closing behavior. One known example 
is the point of $(g,J)=(0,1)$ in the LTC model with 
an irrelevant boundary perturbation $X^{(1)}_{1}Z^{(2)}_{1}Z^{(1)}_{2}+Z^{(2)}_{N-1}Z^{(1)}_{N}X^{(2)}_{N}$ 
which preserves the symmetry generated by the spin-flip and complex conjugation.
The point is shown to reveal an algebraic closing with leading behavior of $1/L^{14}$ 
by second-order perturbation theory of lsing CFT~\cite{verresen2021gapless}. 
Following a common understanding of the universality class, 
it is expected to observe such algebraic closing behavior 
on all the points of the critical line which belongs to the same SEIU class in the bulk. 
However, this is not true as we shall discuss in the following.

We first consider the point $(g,J)=(1,0)$ in SEATU, which is labeled by a black star. By a simple geometric transformation, the ladder model can be kneaded into one chain with cluster coupling and a next-nearest-neighboring (NNN) coupling. The Hamiltonian reads
\be
H_{0} = -\sum^{2N}_{n=1}\left( Z_{n-1} X_{n} Z_{n+1} + gZ_{n-1} Z_{n+1}\right).
\label{Eq:fourde}
\ee
which now preserves the spin flip operator on the odd and even sites: $U_{\text{odd}}=\prod^N_{n=1}X_{2n-1}$ and $U_{\text{even}}=\prod^N_{n=1}X_{2n}$.  To detect the nontrivial edge modes for this criticality, we first write down the critical Hamiltonian in OBC,
\be
H^{\text{OBC}}_{0} = -\sum^{2N-1}_{n=2}\left( Z_{n-1} X_{n} Z_{n+1} + Z_{n-1} Z_{n+1}\right).
\label{Eq:fourde}
\ee
The set of operators $\{Z_1, Z_{2L}, U_{\text{odd}}, U_{\text{even}}\}$ all commute with the Hamiltonian, hence the ground state degeneracy must be at least the dimension of its irreducible representation as four. Applying the $U_{\text{DW}}$ transformation on Eq.~\eqref{Eq:fourde}:
\be
U_{\text{DW}}H^{\text{OBC}}_{0} U^{\dagger}_{\text{DW}}= -\sum^{2N-1}_{n=2}\left( X_{n} + Z_{n-1} Z_{n+1}\right).
\ee
Within the notation of the original triangular ladder, the result is two decoupled critical lsing chains:
\be
\begin{aligned}
U_{\text{DW}}H^{\text{OBC}}_{0} U^{\dagger}_{\text{DW}}= &-\sum^{N}_{n=2}\left( X^{(1)}_{n} + Z^{(1)}_{n-1} Z^{(1)}_{n}\right)\\&-\sum^{N-1}_{n=1}\left( X^{(2)}_{n} + Z^{(2)}_{n} Z^{(2)}_{n+1}\right).
\label{Eq:two decoupled lsing}
\end{aligned}
\ee
The first chain has the boundary condition ($\text{fixed}^{\pm}$, free)
while the second chain has the boundary condition
(free, $\text{fixed}^{\pm}$) \cite{cardy2004boundary,PhysRevB.35.7062}. Here $\text{fixed}^{\pm}$ boundary condition of two chains corresponds to the $Z^{(1)}_1$ and $Z^{(2)}_N$ being $\pm 1$. Thus the ground states should have a four-fold degeneracy with corresponding spontaneous boundary magnetization.
We can add a symmetric marginal boundary perturbation $\lambda_{1}X_{1} + \lambda_{2}X_{2N}$ which lifts this degeneracy and leads to the opening of a gap. Such boundary term does not commute with $Z_{2L}$ and $Z_{1}$, thus the set of operators commuting with the Hamiltonian $H_{0}^{\text{OBC}}+\lambda_{1}X_{1} + \lambda_{2}X_{2N}$ reduce to $\{U_{\text{odd}}, U_{\text{even}}\}$. As a consequence, the dimension of irreducible representation reduces from four to one \cite{fn_lambda}. Such finite size energy gap shows algebraic closing behaviour as the system size increases. By setting both $\lambda_{1}$ and $\lambda_{2}$ as small values, the leading term of the finite size energy splitting between the ground state and the first excited state $\Delta E = E_{1}-E_{0}$ scales as $1/L$~\cite{SM}, while higher order correction to this behaviour  becomes necessary for larger $\lambda_{1}$ and $\lambda_{2}$.  Additionally,  when considering the system with a fixed length, the energy gap $\Delta E$ follows linearly with both $\lambda_{1}$ and $\lambda_{2}$. The result is shown in Fig.~\ref{fig2}.

Besides the special SEATU point, it is expected that all 
other points along the critical line, which belong to SEIU class, should follow an algebraic behaviour of $1/L^{14}$  with an irrelevant perturbation $X^{(1)}_{1}Z^{(2)}_{1}Z^{(1)}_{2}+Z^{(2)}_{N-1}Z^{(1)}_{N}X^{(2)}_{N}$~\cite{verresen2021gapless}. However, after we exhaust the whole critical line, we find that except the diamond point $(g,J)=(0,1)$ 
with known finite-size gap closing behaviour as $1/L^{14}$~\cite{verresen2021gapless}, the other points host an exponential closing behaviour  as $\exp{(-L/\xi_{loc})}$ (Seeing Fig.~\ref{fig3} (a) and (b) for two instances with $J=0.4$ and $J=1.4$).
For further characterizing the detailed gap closing behaviour along this critical line, 
we define the corresponding constant $\xi_{loc}$ as the edge mode localization length. 
The algebraic closing behaviour corresponds to a divergence of $\xi_{loc}\rightarrow\infty$. 
Comparing the result of the correlation length defined by the static two point correlation function $\langle O_j O_i\rangle\sim\exp{(-|i-j|/\xi)}$ (we call it as the bulk correlation length in the following), 
we find that $\xi_{loc}$ goes to the same limit as $\xi$ when the system is tuned
away from the critical point. But if considering the exponential closing points 
on the critical line, $\xi$ diverges near the critical point while $\xi_{loc}$ does not (Seeing Fig.~\ref{fig3} (c) for instance with $J=0.4$). 
This indicates that the $\xi_{loc}$ shows a  fine and rich  structure of a universality class which cannot be described by $\xi$ and the smooth path preserving symmetry. 

\textit{Conclusion and discussion.---}In this letter, we propose the spin LTC model and identify the symmetry-enriched critical line that separates the SPT phase and the SSB phase. The universality classes
along the critical line are exhausted as $(g,J)=(1,0)$ falls into the SEATU, 
and all the other points fall into the SEIU. The presence of 
degenerate edge modes under the OBC is a characteristic of the SEQC. We further investigate the stability of such edge modes by adding symmetric boundary perturbations. For the SEATU at $(g,J)=(1,0)$,
a marginal perturbation  results in a novel algebraic energy splitting closing behaviour scaling as $1/L$. For the SEIU, an irrelevant perturbation induces algebraic energy gap closing behaviour $1/L^{14}$ at the special
point $(g,J)=(0,1)$ and exponential energy gap closing behaviour $\exp{(-L/\xi_{loc})}$ at all the other points. By defining the corresponding constant $\xi_{loc}$ as the edge mode localization length, we find that it shows different asymptotic behaviours from the bulk correlation length (defined by the correlation function) near the critical points.
Since $\xi_{loc}$ corresponds to the involvement of additional gapped 
sectors with gap $\Delta E \sim 1/\xi_{loc}$, it cannot be distinguished by low energy theory. 
We propose that the special point $(g,J)=(0,1)$ with $\xi_{loc}\to \infty$ implies the fine and rich structure of SEQC, demonstrating characteristics that go beyond the previous understanding of SEQC, and challenge the traditional paradigm of phase transition.

\begin{figure*}
	\includegraphics[width=17cm]{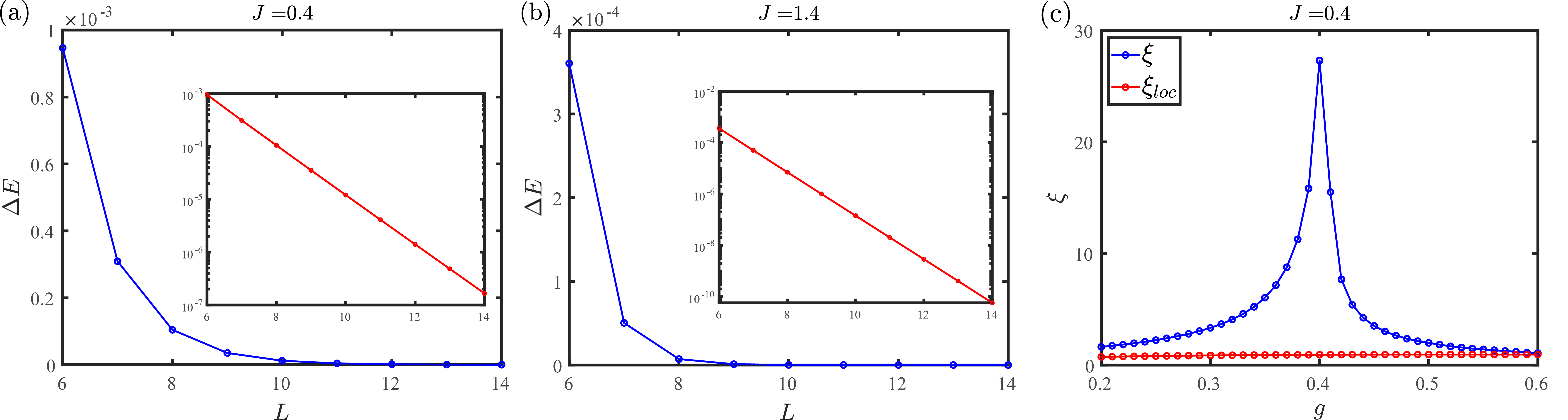}
	\caption{(a) and (b) The gap closing behaviour for $J=0.4$ and $J=1.4$ on the critical line with system size $L$ being from 6 to 14, respectively. Inset: The semi-log plot reveals the exponential gap closing behaviour. (c) The edge mode localization length $\xi_{loc}$ and bulk correlation length $\xi$ as a function of $g$ for $J=0.4$. The bulk correlation length is calculated by $\langle Z_{j}^{(1)}Z_{i}^{(1)}\rangle\sim \exp{(-|j-i|/\xi)}$.}
	\label{fig3}
\end{figure*}

\textit{Acknowledgments.---}We thank Ning Xi for useful discussions.
The work at Shanghai Jiao Tong University is sponsored by
National Natural Science Foundation of China No. 12274288 and the Innovation Program for Quantum Science and Technology Grant No. 2021ZD0301900 and
Natural Science Foundation of Shanghai with Grant No. 20ZR1428400.
L.L. acknowledges support from Global Science Graduate Course (GSGC) program at the University of Tokyo.

\bibliography{bib}

\begin{thebibliography}{47}
\expandafter\ifx\csname natexlab\endcsname\relax\def\natexlab#1{#1}\fi
\expandafter\ifx\csname bibnamefont\endcsname\relax
  \def\bibnamefont#1{#1}\fi
\expandafter\ifx\csname bibfnamefont\endcsname\relax
  \def\bibfnamefont#1{#1}\fi
\expandafter\ifx\csname citenamefont\endcsname\relax
  \def\citenamefont#1{#1}\fi
\expandafter\ifx\csname url\endcsname\relax
  \def\url#1{\texttt{#1}}\fi
\expandafter\ifx\csname urlprefix\endcsname\relax\def\urlprefix{URL }\fi
\providecommand{\bibinfo}[2]{#2}
\providecommand{\eprint}[2][]{\url{#2}}

\bibitem[{\citenamefont{Chen et~al.}(2010)\citenamefont{Chen, Gu, and
  Wen}}]{PhysRevB.82.155138}
\bibinfo{author}{\bibfnamefont{X.}~\bibnamefont{Chen}},
  \bibinfo{author}{\bibfnamefont{Z.-C.} \bibnamefont{Gu}}, \bibnamefont{and}
  \bibinfo{author}{\bibfnamefont{X.-G.} \bibnamefont{Wen}},
  \bibinfo{journal}{Phys. Rev. B} \textbf{\bibinfo{volume}{82}},
  \bibinfo{pages}{155138} (\bibinfo{year}{2010}).

\bibitem[{\citenamefont{Levin and Gu}(2012)}]{PhysRevB.86.115109}
\bibinfo{author}{\bibfnamefont{M.}~\bibnamefont{Levin}} \bibnamefont{and}
  \bibinfo{author}{\bibfnamefont{Z.-C.} \bibnamefont{Gu}},
  \bibinfo{journal}{Phys. Rev. B} \textbf{\bibinfo{volume}{86}},
  \bibinfo{pages}{115109} (\bibinfo{year}{2012}).

\bibitem[{\citenamefont{Wang et~al.}(2015)\citenamefont{Wang, Gu, and
  Wen}}]{PhysRevLett.114.031601}
\bibinfo{author}{\bibfnamefont{J.~C.} \bibnamefont{Wang}},
  \bibinfo{author}{\bibfnamefont{Z.-C.} \bibnamefont{Gu}}, \bibnamefont{and}
  \bibinfo{author}{\bibfnamefont{X.-G.} \bibnamefont{Wen}},
  \bibinfo{journal}{Phys. Rev. Lett.} \textbf{\bibinfo{volume}{114}},
  \bibinfo{pages}{031601} (\bibinfo{year}{2015}).

\bibitem[{\citenamefont{Chen et~al.}(2013{\natexlab{a}})\citenamefont{Chen, Gu,
  Liu, and Wen}}]{PhysRevB.87.155114}
\bibinfo{author}{\bibfnamefont{X.}~\bibnamefont{Chen}},
  \bibinfo{author}{\bibfnamefont{Z.-C.} \bibnamefont{Gu}},
  \bibinfo{author}{\bibfnamefont{Z.-X.} \bibnamefont{Liu}}, \bibnamefont{and}
  \bibinfo{author}{\bibfnamefont{X.-G.} \bibnamefont{Wen}},
  \bibinfo{journal}{Phys. Rev. B} \textbf{\bibinfo{volume}{87}},
  \bibinfo{pages}{155114} (\bibinfo{year}{2013}{\natexlab{a}}).

\bibitem[{\citenamefont{Chen et~al.}(2012)\citenamefont{Chen, Gu, Liu, and
  Wen}}]{chen2012symmetry}
\bibinfo{author}{\bibfnamefont{X.}~\bibnamefont{Chen}},
  \bibinfo{author}{\bibfnamefont{Z.-C.} \bibnamefont{Gu}},
  \bibinfo{author}{\bibfnamefont{Z.-X.} \bibnamefont{Liu}}, \bibnamefont{and}
  \bibinfo{author}{\bibfnamefont{X.-G.} \bibnamefont{Wen}},
  \bibinfo{journal}{Science} \textbf{\bibinfo{volume}{338}},
  \bibinfo{pages}{1604} (\bibinfo{year}{2012}).

\bibitem[{\citenamefont{Else and Nayak}(2014)}]{PhysRevB.90.235137}
\bibinfo{author}{\bibfnamefont{D.~V.} \bibnamefont{Else}} \bibnamefont{and}
  \bibinfo{author}{\bibfnamefont{C.}~\bibnamefont{Nayak}},
  \bibinfo{journal}{Phys. Rev. B} \textbf{\bibinfo{volume}{90}},
  \bibinfo{pages}{235137} (\bibinfo{year}{2014}).

\bibitem[{\citenamefont{Wen}(2014)}]{PhysRevB.89.035147}
\bibinfo{author}{\bibfnamefont{X.-G.} \bibnamefont{Wen}},
  \bibinfo{journal}{Phys. Rev. B} \textbf{\bibinfo{volume}{89}},
  \bibinfo{pages}{035147} (\bibinfo{year}{2014}).

\bibitem[{\citenamefont{Schuch et~al.}(2011)\citenamefont{Schuch,
  P\'erez-Garc\'{\i}a, and Cirac}}]{PhysRevB.84.165139}
\bibinfo{author}{\bibfnamefont{N.}~\bibnamefont{Schuch}},
  \bibinfo{author}{\bibfnamefont{D.}~\bibnamefont{P\'erez-Garc\'{\i}a}},
  \bibnamefont{and} \bibinfo{author}{\bibfnamefont{I.}~\bibnamefont{Cirac}},
  \bibinfo{journal}{Phys. Rev. B} \textbf{\bibinfo{volume}{84}},
  \bibinfo{pages}{165139} (\bibinfo{year}{2011}).

\bibitem[{\citenamefont{Affleck et~al.}(1987)\citenamefont{Affleck, Kennedy,
  Lieb, and Tasaki}}]{PhysRevLett.59.799}
\bibinfo{author}{\bibfnamefont{I.}~\bibnamefont{Affleck}},
  \bibinfo{author}{\bibfnamefont{T.}~\bibnamefont{Kennedy}},
  \bibinfo{author}{\bibfnamefont{E.~H.} \bibnamefont{Lieb}}, \bibnamefont{and}
  \bibinfo{author}{\bibfnamefont{H.}~\bibnamefont{Tasaki}},
  \bibinfo{journal}{Phys. Rev. Lett.} \textbf{\bibinfo{volume}{59}},
  \bibinfo{pages}{799} (\bibinfo{year}{1987}).

\bibitem[{\citenamefont{Chen et~al.}(2011)\citenamefont{Chen, Gu, and
  Wen}}]{PhysRevB.83.035107}
\bibinfo{author}{\bibfnamefont{X.}~\bibnamefont{Chen}},
  \bibinfo{author}{\bibfnamefont{Z.-C.} \bibnamefont{Gu}}, \bibnamefont{and}
  \bibinfo{author}{\bibfnamefont{X.-G.} \bibnamefont{Wen}},
  \bibinfo{journal}{Phys. Rev. B} \textbf{\bibinfo{volume}{83}},
  \bibinfo{pages}{035107} (\bibinfo{year}{2011}).

\bibitem[{\citenamefont{Pollmann et~al.}(2010)\citenamefont{Pollmann, Turner,
  Berg, and Oshikawa}}]{PhysRevB.81.064439}
\bibinfo{author}{\bibfnamefont{F.}~\bibnamefont{Pollmann}},
  \bibinfo{author}{\bibfnamefont{A.~M.} \bibnamefont{Turner}},
  \bibinfo{author}{\bibfnamefont{E.}~\bibnamefont{Berg}}, \bibnamefont{and}
  \bibinfo{author}{\bibfnamefont{M.}~\bibnamefont{Oshikawa}},
  \bibinfo{journal}{Phys. Rev. B} \textbf{\bibinfo{volume}{81}},
  \bibinfo{pages}{064439} (\bibinfo{year}{2010}).

\bibitem[{\citenamefont{Pollmann et~al.}(2012)\citenamefont{Pollmann, Berg,
  Turner, and Oshikawa}}]{PhysRevB.85.075125}
\bibinfo{author}{\bibfnamefont{F.}~\bibnamefont{Pollmann}},
  \bibinfo{author}{\bibfnamefont{E.}~\bibnamefont{Berg}},
  \bibinfo{author}{\bibfnamefont{A.~M.} \bibnamefont{Turner}},
  \bibnamefont{and} \bibinfo{author}{\bibfnamefont{M.}~\bibnamefont{Oshikawa}},
  \bibinfo{journal}{Phys. Rev. B} \textbf{\bibinfo{volume}{85}},
  \bibinfo{pages}{075125} (\bibinfo{year}{2012}).

\bibitem[{\citenamefont{Sachdev}(2011)}]{sachdev_2011}
\bibinfo{author}{\bibfnamefont{S.}~\bibnamefont{Sachdev}}, {\it
  \bibinfo{title}{Quantum Phase Transitions}} (\bibinfo{publisher}{Cambridge
  University Press}, \bibinfo{address}{Cambridge, England},
  \bibinfo{year}{2011}).

\bibitem[{\citenamefont{Verresen
  et~al.}(2021{\natexlab{a}})\citenamefont{Verresen, Thorngren, Jones, and
  Pollmann}}]{verresen2021gapless}
\bibinfo{author}{\bibfnamefont{R.}~\bibnamefont{Verresen}},
  \bibinfo{author}{\bibfnamefont{R.}~\bibnamefont{Thorngren}},
  \bibinfo{author}{\bibfnamefont{N.~G.} \bibnamefont{Jones}}, \bibnamefont{and}
  \bibinfo{author}{\bibfnamefont{F.}~\bibnamefont{Pollmann}},
  \bibinfo{journal}{Physical Review X} \textbf{\bibinfo{volume}{11}},
  \bibinfo{pages}{041059} (\bibinfo{year}{2021}{\natexlab{a}}).

\bibitem[{\citenamefont{Yu et~al.}(2022)\citenamefont{Yu, Huang, Song, Xu,
  Ding, and Zhang}}]{longzhang_2022}
\bibinfo{author}{\bibfnamefont{X.-J.} \bibnamefont{Yu}},
  \bibinfo{author}{\bibfnamefont{R.-Z.} \bibnamefont{Huang}},
  \bibinfo{author}{\bibfnamefont{H.-H.} \bibnamefont{Song}},
  \bibinfo{author}{\bibfnamefont{L.}~\bibnamefont{Xu}},
  \bibinfo{author}{\bibfnamefont{C.}~\bibnamefont{Ding}}, \bibnamefont{and}
  \bibinfo{author}{\bibfnamefont{L.}~\bibnamefont{Zhang}},
  \bibinfo{journal}{Phys. Rev. Lett.} \textbf{\bibinfo{volume}{129}},
  \bibinfo{pages}{210601} (\bibinfo{year}{2022}).

\bibitem[{fn_({\natexlab{a}})}]{fn_smoothpath}
\bibinfo{howpublished}{After discretization, such a smooth path is equivalent
  to locally-symmetric unitary transformation, which is well-known in quantum
  information.}

\bibitem[{\citenamefont{Scaffidi et~al.}(2017)\citenamefont{Scaffidi, Parker,
  and Vasseur}}]{PhysRevX.7.041048}
\bibinfo{author}{\bibfnamefont{T.}~\bibnamefont{Scaffidi}},
  \bibinfo{author}{\bibfnamefont{D.~E.} \bibnamefont{Parker}},
  \bibnamefont{and} \bibinfo{author}{\bibfnamefont{R.}~\bibnamefont{Vasseur}},
  \bibinfo{journal}{Phys. Rev. X} \textbf{\bibinfo{volume}{7}},
  \bibinfo{pages}{041048} (\bibinfo{year}{2017}).

\bibitem[{\citenamefont{Verresen
  et~al.}(2021{\natexlab{b}})\citenamefont{Verresen, Thorngren, Jones, and
  Pollmann}}]{PhysRevX.11.041059}
\bibinfo{author}{\bibfnamefont{R.}~\bibnamefont{Verresen}},
  \bibinfo{author}{\bibfnamefont{R.}~\bibnamefont{Thorngren}},
  \bibinfo{author}{\bibfnamefont{N.~G.} \bibnamefont{Jones}}, \bibnamefont{and}
  \bibinfo{author}{\bibfnamefont{F.}~\bibnamefont{Pollmann}},
  \bibinfo{journal}{Phys. Rev. X} \textbf{\bibinfo{volume}{11}},
  \bibinfo{pages}{041059} (\bibinfo{year}{2021}{\natexlab{b}}).

\bibitem[{\citenamefont{Parker et~al.}(2018)\citenamefont{Parker, Scaffidi, and
  Vasseur}}]{PhysRevB.97.165114}
\bibinfo{author}{\bibfnamefont{D.~E.} \bibnamefont{Parker}},
  \bibinfo{author}{\bibfnamefont{T.}~\bibnamefont{Scaffidi}}, \bibnamefont{and}
  \bibinfo{author}{\bibfnamefont{R.}~\bibnamefont{Vasseur}},
  \bibinfo{journal}{Phys. Rev. B} \textbf{\bibinfo{volume}{97}},
  \bibinfo{pages}{165114} (\bibinfo{year}{2018}).

\bibitem[{\citenamefont{Thorngren et~al.}(2021)\citenamefont{Thorngren,
  Vishwanath, and Verresen}}]{PhysRevB.104.075132}
\bibinfo{author}{\bibfnamefont{R.}~\bibnamefont{Thorngren}},
  \bibinfo{author}{\bibfnamefont{A.}~\bibnamefont{Vishwanath}},
  \bibnamefont{and} \bibinfo{author}{\bibfnamefont{R.}~\bibnamefont{Verresen}},
  \bibinfo{journal}{Phys. Rev. B} \textbf{\bibinfo{volume}{104}},
  \bibinfo{pages}{075132} (\bibinfo{year}{2021}).

\bibitem[{\citenamefont{Parker et~al.}(2019)\citenamefont{Parker, Vasseur, and
  Scaffidi}}]{PhysRevLett.122.240605}
\bibinfo{author}{\bibfnamefont{D.~E.} \bibnamefont{Parker}},
  \bibinfo{author}{\bibfnamefont{R.}~\bibnamefont{Vasseur}}, \bibnamefont{and}
  \bibinfo{author}{\bibfnamefont{T.}~\bibnamefont{Scaffidi}},
  \bibinfo{journal}{Phys. Rev. Lett.} \textbf{\bibinfo{volume}{122}},
  \bibinfo{pages}{240605} (\bibinfo{year}{2019}).

\bibitem[{\citenamefont{Yang et~al.}(2022)\citenamefont{Yang, Li, Okunishi, and
  Katsura}}]{yang2022duality}
\bibinfo{author}{\bibfnamefont{H.}~\bibnamefont{Yang}},
  \bibinfo{author}{\bibfnamefont{L.}~\bibnamefont{Li}},
  \bibinfo{author}{\bibfnamefont{K.}~\bibnamefont{Okunishi}}, \bibnamefont{and}
  \bibinfo{author}{\bibfnamefont{H.}~\bibnamefont{Katsura}},
  \bibinfo{journal}{arXiv preprint arXiv:2203.15791}  (\bibinfo{year}{2022}).

\bibitem[{\citenamefont{Li et~al.}(2022)\citenamefont{Li, Oshikawa, and
  Zheng}}]{li2022symmetry}
\bibinfo{author}{\bibfnamefont{L.}~\bibnamefont{Li}},
  \bibinfo{author}{\bibfnamefont{M.}~\bibnamefont{Oshikawa}}, \bibnamefont{and}
  \bibinfo{author}{\bibfnamefont{Y.}~\bibnamefont{Zheng}},
  \bibinfo{journal}{arXiv preprint arXiv:2204.03131}  (\bibinfo{year}{2022}).

\bibitem[{\citenamefont{Li et~al.}(2023)\citenamefont{Li, Oshikawa, and
  Zheng}}]{Li:2023mmw}
\bibinfo{author}{\bibfnamefont{L.}~\bibnamefont{Li}},
  \bibinfo{author}{\bibfnamefont{M.}~\bibnamefont{Oshikawa}}, \bibnamefont{and}
  \bibinfo{author}{\bibfnamefont{Y.}~\bibnamefont{Zheng}},
  \bibinfo{journal}{arXiv preprint arXiv:2301.07899}  (\bibinfo{year}{2023}).

\bibitem[{\citenamefont{Hidaka et~al.}(2022)\citenamefont{Hidaka, Furuya, Ueda,
  and Tada}}]{PhysRevB.106.144436}
\bibinfo{author}{\bibfnamefont{Y.}~\bibnamefont{Hidaka}},
  \bibinfo{author}{\bibfnamefont{S.~C.} \bibnamefont{Furuya}},
  \bibinfo{author}{\bibfnamefont{A.}~\bibnamefont{Ueda}}, \bibnamefont{and}
  \bibinfo{author}{\bibfnamefont{Y.}~\bibnamefont{Tada}},
  \bibinfo{journal}{Phys. Rev. B} \textbf{\bibinfo{volume}{106}},
  \bibinfo{pages}{144436} (\bibinfo{year}{2022}).

\bibitem[{\citenamefont{Vidal}(2003)}]{vidal_tebd}
\bibinfo{author}{\bibfnamefont{G.}~\bibnamefont{Vidal}},
  \bibinfo{journal}{Phys. Rev. Lett.} \textbf{\bibinfo{volume}{91}},
  \bibinfo{pages}{147902} (\bibinfo{year}{2003}).

\bibitem[{\citenamefont{Oshikawa and Affleck}(1997)}]{OSHIKAWA1997533}
\bibinfo{author}{\bibfnamefont{M.}~\bibnamefont{Oshikawa}} \bibnamefont{and}
  \bibinfo{author}{\bibfnamefont{I.}~\bibnamefont{Affleck}},
  \bibinfo{journal}{Nuclear Physics B} \textbf{\bibinfo{volume}{495}},
  \bibinfo{pages}{533} (\bibinfo{year}{1997}), ISSN \bibinfo{issn}{0550-3213}.

\bibitem[{\citenamefont{Oshikawa and Affleck}(1996)}]{PhysRevLett.77.2604}
\bibinfo{author}{\bibfnamefont{M.}~\bibnamefont{Oshikawa}} \bibnamefont{and}
  \bibinfo{author}{\bibfnamefont{I.}~\bibnamefont{Affleck}},
  \bibinfo{journal}{Phys. Rev. Lett.} \textbf{\bibinfo{volume}{77}},
  \bibinfo{pages}{2604} (\bibinfo{year}{1996}).

\bibitem[{\citenamefont{Ginsparg}(1988)}]{ginsparg1988applied}
\bibinfo{author}{\bibfnamefont{P.}~\bibnamefont{Ginsparg}},
  \bibinfo{journal}{arXiv preprint hep-th/9108028}  (\bibinfo{year}{1988}).

\bibitem[{\citenamefont{Amit et~al.}(2010)\citenamefont{Amit, Gabriel, Bikas,
  Uma, Thomas, Rosenbaum, and Diptiman}}]{Ising_review}
\bibinfo{author}{\bibfnamefont{D.}~\bibnamefont{Amit}},
  \bibinfo{author}{\bibfnamefont{A.}~\bibnamefont{Gabriel}},
  \bibinfo{author}{\bibfnamefont{K.~C.} \bibnamefont{Bikas}},
  \bibinfo{author}{\bibfnamefont{D.}~\bibnamefont{Uma}},
  \bibinfo{author}{\bibfnamefont{F.}~\bibnamefont{Thomas}},
  \bibinfo{author}{\bibnamefont{Rosenbaum}}, \bibnamefont{and}
  \bibinfo{author}{\bibfnamefont{S.}~\bibnamefont{Diptiman}},
  \bibinfo{journal}{arXiv:1012.0653}  (\bibinfo{year}{2010}).

\bibitem[{\citenamefont{Chen et~al.}(2013{\natexlab{b}})\citenamefont{Chen, Lu,
  and Vishwanath}}]{Chen2013SymmetryprotectedTP}
\bibinfo{author}{\bibfnamefont{X.}~\bibnamefont{Chen}},
  \bibinfo{author}{\bibfnamefont{Y.-M.} \bibnamefont{Lu}}, \bibnamefont{and}
  \bibinfo{author}{\bibfnamefont{A.}~\bibnamefont{Vishwanath}},
  \bibinfo{journal}{Nature communications} \textbf{\bibinfo{volume}{5}},
  \bibinfo{pages}{3507} (\bibinfo{year}{2013}{\natexlab{b}}).

\bibitem[{\citenamefont{Haldane}(1983)}]{PhysRevLett.50.1153}
\bibinfo{author}{\bibfnamefont{F.~D.~M.} \bibnamefont{Haldane}},
  \bibinfo{journal}{Phys. Rev. Lett.} \textbf{\bibinfo{volume}{50}},
  \bibinfo{pages}{1153} (\bibinfo{year}{1983}).

\bibitem[{\citenamefont{den Nijs and Rommelse}(1989)}]{PhysRevB.40.4709}
\bibinfo{author}{\bibfnamefont{M.}~\bibnamefont{den Nijs}} \bibnamefont{and}
  \bibinfo{author}{\bibfnamefont{K.}~\bibnamefont{Rommelse}},
  \bibinfo{journal}{Phys. Rev. B} \textbf{\bibinfo{volume}{40}},
  \bibinfo{pages}{4709} (\bibinfo{year}{1989}).

\bibitem[{\citenamefont{Santos}(2015)}]{PhysRevB.91.155150}
\bibinfo{author}{\bibfnamefont{L.~H.} \bibnamefont{Santos}},
  \bibinfo{journal}{Phys. Rev. B} \textbf{\bibinfo{volume}{91}},
  \bibinfo{pages}{155150} (\bibinfo{year}{2015}).

\bibitem[{\citenamefont{Li and Yao}(2022)}]{PhysRevB.106.224420}
\bibinfo{author}{\bibfnamefont{L.}~\bibnamefont{Li}} \bibnamefont{and}
  \bibinfo{author}{\bibfnamefont{Y.}~\bibnamefont{Yao}},
  \bibinfo{journal}{Phys. Rev. B} \textbf{\bibinfo{volume}{106}},
  \bibinfo{pages}{224420} (\bibinfo{year}{2022}).

\bibitem[{\citenamefont{Ramos et~al.}(2020)\citenamefont{Ramos, Lencs\'es,
  Xavier, and Pereira}}]{ladderTFIC}
\bibinfo{author}{\bibfnamefont{F.~B.} \bibnamefont{Ramos}},
  \bibinfo{author}{\bibfnamefont{M.}~\bibnamefont{Lencs\'es}},
  \bibinfo{author}{\bibfnamefont{J.~C.} \bibnamefont{Xavier}},
  \bibnamefont{and} \bibinfo{author}{\bibfnamefont{R.~G.}
  \bibnamefont{Pereira}}, \bibinfo{journal}{Phys. Rev. B}
  \textbf{\bibinfo{volume}{102}}, \bibinfo{pages}{014426}
  (\bibinfo{year}{2020}).

\bibitem[{\citenamefont{Boyanovsky}(1989)}]{IsingCFT1989}
\bibinfo{author}{\bibfnamefont{D.}~\bibnamefont{Boyanovsky}},
  \bibinfo{journal}{Phys. Rev. B} \textbf{\bibinfo{volume}{39}},
  \bibinfo{pages}{6744} (\bibinfo{year}{1989}).

\bibitem[{\citenamefont{LeClair et~al.}(1998)\citenamefont{LeClair, Ludwig, and
  Mussardo}}]{LECLAIR1998523}
\bibinfo{author}{\bibfnamefont{A.}~\bibnamefont{LeClair}},
  \bibinfo{author}{\bibfnamefont{A.}~\bibnamefont{Ludwig}}, \bibnamefont{and}
  \bibinfo{author}{\bibfnamefont{G.}~\bibnamefont{Mussardo}},
  \bibinfo{journal}{Nuclear Physics B} \textbf{\bibinfo{volume}{512}},
  \bibinfo{pages}{523} (\bibinfo{year}{1998}), ISSN \bibinfo{issn}{0550-3213}.

\bibitem[{\citenamefont{Banks et~al.}(1976)\citenamefont{Banks, Horn, and
  Neuberger}}]{BANKS1976119}
\bibinfo{author}{\bibfnamefont{T.}~\bibnamefont{Banks}},
  \bibinfo{author}{\bibfnamefont{D.}~\bibnamefont{Horn}}, \bibnamefont{and}
  \bibinfo{author}{\bibfnamefont{H.}~\bibnamefont{Neuberger}},
  \bibinfo{journal}{Nuclear Physics B} \textbf{\bibinfo{volume}{108}},
  \bibinfo{pages}{119} (\bibinfo{year}{1976}), ISSN \bibinfo{issn}{0550-3213}.

\bibitem[{\citenamefont{Cardy}(2004)}]{cardy2004boundary}
\bibinfo{author}{\bibfnamefont{J.}~\bibnamefont{Cardy}},
  \bibinfo{journal}{arXiv preprint hep-th/0411189}  (\bibinfo{year}{2004}).

\bibitem[{\citenamefont{Cabrera and Jullien}(1987)}]{PhysRevB.35.7062}
\bibinfo{author}{\bibfnamefont{G.~G.} \bibnamefont{Cabrera}} \bibnamefont{and}
  \bibinfo{author}{\bibfnamefont{R.}~\bibnamefont{Jullien}},
  \bibinfo{journal}{Phys. Rev. B} \textbf{\bibinfo{volume}{35}},
  \bibinfo{pages}{7062} (\bibinfo{year}{1987}).

\bibitem[{fn_({\natexlab{b}})}]{fn_lambda}
\bibinfo{howpublished}{It is straightforward to check that only both
  $\lambda_{1}$ and $\lambda_{2}$ is non-zero the degeneracy could be broken,
  otherwise there will be at least two-fold degeneracy left.}

\bibitem[{SM()}]{SM}
\bibinfo{howpublished}{See Supplemental Material at [URL will be inserted by
  publisher] for detailed calculation and lengthy results.}

\bibitem[{\citenamefont{Watts}(2001)}]{watts2001boundary}
\bibinfo{author}{\bibfnamefont{G.}~\bibnamefont{Watts}},
  \bibinfo{journal}{Nuclear Physics B} \textbf{\bibinfo{volume}{596}},
  \bibinfo{pages}{513} (\bibinfo{year}{2001}).

\bibitem[{fn_({\natexlab{c}})}]{fn_kw}
\bibinfo{howpublished}{One can directly show this result by KW duality, as it
  exchanges free and fixed boundary condition, also exchanges the boundary $X$
  and $Z$ operator at the same time.}

\bibitem[{\citenamefont{Barouch and McCoy}(1971)}]{Mccoy_1971}
\bibinfo{author}{\bibfnamefont{E.}~\bibnamefont{Barouch}} \bibnamefont{and}
  \bibinfo{author}{\bibfnamefont{B.~M.} \bibnamefont{McCoy}},
  \bibinfo{journal}{Phys. Rev. A} \textbf{\bibinfo{volume}{3}},
  \bibinfo{pages}{786} (\bibinfo{year}{1971}).

\bibitem[{fn_({\natexlab{d}})}]{fn_lambda_supp}
\bibinfo{howpublished}{When $\lambda_1=1$ and $\lambda_2$ is small, the
  spectrum is the same.}

\end{thebibliography}
\bibliographystyle{apsrev-nourl}

\newpage

\appendix

\onecolumngrid

\section*{{ \Large Supplementary Material}}

\section{The algebraic decaying behaviour of the gap closing for SEATU}
Here we provide an analytical approach to demonstrate the $1/L$ scaling behaviour when $\lambda_1=\lambda_2$. The Hamiltonian after $U_{DW}$ is
\be
\begin{aligned}
H_{1}=&U_{DW}(H^{\text{OBC}}_{0}+\lambda_1 X_1+\lambda X_{2N}) U^{\dagger}_{DW}\\= &-\sum^{N}_{n=2}\left( X^{(1)}_{n} + Z^{(1)}_{n-1} Z^{(1)}_{n}\right)-\lambda_1 X^{(1)}_1 Z^{(2)}_{1}-\sum^{N-1}_{n=1}\left( X^{(2)}_{n} + Z^{(2)}_{n} Z^{(2)}_{n+1}\right)-\lambda_2 X^{(2)}_N Z^{(1)}_{N}.
\end{aligned}\label{Eq:perted Hal}
\ee
In the low energy, $Z^{(2)}_{1}$ and $Z^{(1)}_{N}$ correspond to the boundary operator $\sigma^{(2)}(x=0)$ and $\sigma^{(1)}(x=L)$. Since the second chain has free boundary condition at left side and the first chain has free boundary condition at right side, the boundary scaling dimension of $Z^{(2)}_{1}$ and $Z^{(1)}_{N}$ are both $\frac{1}{2}$ \cite{watts2001boundary}. Moreover, $X^{1}_{1}$ and $X^{2}_{N}$ correspond to the boundary operator $\mu^{(1)}(x=0)$ and $\mu^{(2)}(x=L)$. Similarly,  since the second chain has fixed boundary condition at left side and the first chain has fixed boundary condition at right side, the boundary scaling dimension of $X^{1}_{1}$ and $X^{2}_{N}$ are also both $\frac{1}{2}$ \footnote{One can directly show this result by KW duality, as it exchanges free and fixed boundary condition, also exchanges the boundary $X$ and $Z$ operator at the same time.}

We confirm this feature by using a Bogouliubov de-Gennes (BdG) formalism for transverse field Ising chain with free OBC on both sides~\cite{Mccoy_1971}, and the result is shown in Fig.~\ref{fig3} . The two-point correlation function $\langle Z_{1}Z_{L}\rangle=\langle\sigma_{1}^{z}\sigma_{L}^{z}\rangle$ is calculated, and the asymptotic behavior
\be
\langle\sigma_{1}^{z}\sigma_{L}^{z}\rangle \sim L^{2\Delta}
\label{Eq:scaling}
\ee
is verified with a very high accuracy as $2\Delta = -0.9961(6)$ for $L$ going from 80 to 200, implying the boundary scaling dimension of $[Z_{\partial}]$ being $1/2$, as the same as the result given by boundary CFT.
Therefore, the perturbation in the Eq.\eqref{Eq:perted Hal} has the scaling dimension 1 which is marginal in the 0+1d RG. Then we can obtain that the scaling dimension of $\lambda_{1/2}$ is 0. Since the $X^{(1)}_1 Z^{(2)}_{1}$ and $Z^{(1)}_N X^{(2)}_{N}$ flip boundary magnetization on the left side and right side respectively, the first order perturbation is enough to observe a splitting:
$\Delta E_{(1)}\propto \frac{\lambda}{L^{\beta}}$. As it has units of energy, we obtain that $\beta=1$. In fact, since $\lambda$ is dimensionless, any order perturbation energy is $\Delta E_{(k)}\propto \frac{\lambda^{n}}{L^1}$. This can explain why our numerical result fits the $1/L$ behaviour very well.

Now let's consider the second case where $\lambda_2=1$ and $\lambda_1$ is small without loss of generality \footnote{When $\lambda_1=1$ and $\lambda_2$ is small, the spectrum is the same.}. We can apply the KW transformation on the second lsing critical chain
\begin{equation}
\begin{aligned}
  &  V_{\text{KW}} Z^{(2)}_{n} V_{\text{KW}}^\dagger = \prod_{k=1}^{n} X^{(2)}_k,\\  &V_{\text{KW}} X^{(2)}_n V_{\text{KW}}^\dagger =
    \begin{cases}
    Z^{(2)}_{n} Z^{(2)}_{n+1}, & n=1, ..., N-1\\
    Z^{(2)}_{N},& n=N
    \end{cases}
    \end{aligned}
\end{equation}
The Hamiltonian after KW transformation is
\be
\begin{aligned}
V_{\text{KW}}H_1 V_{\text{KW}}^\dagger= -\sum^{N}_{n=2}\left( X^{(1)}_{n} + Z^{(1)}_{n-1} Z^{(1)}_{n}\right)-\lambda_1 X^{(1)}_1 X^{(2)}_1-\sum^{N}_{n=2}\left( X^{(2)}_{n} + Z^{(2)}_{n-1} Z^{(2)}_{n}\right)-Z^{(1)}_{N}Z^{(2)}_N.
\end{aligned}
\ee

\begin{figure}
	\includegraphics[width=8cm]{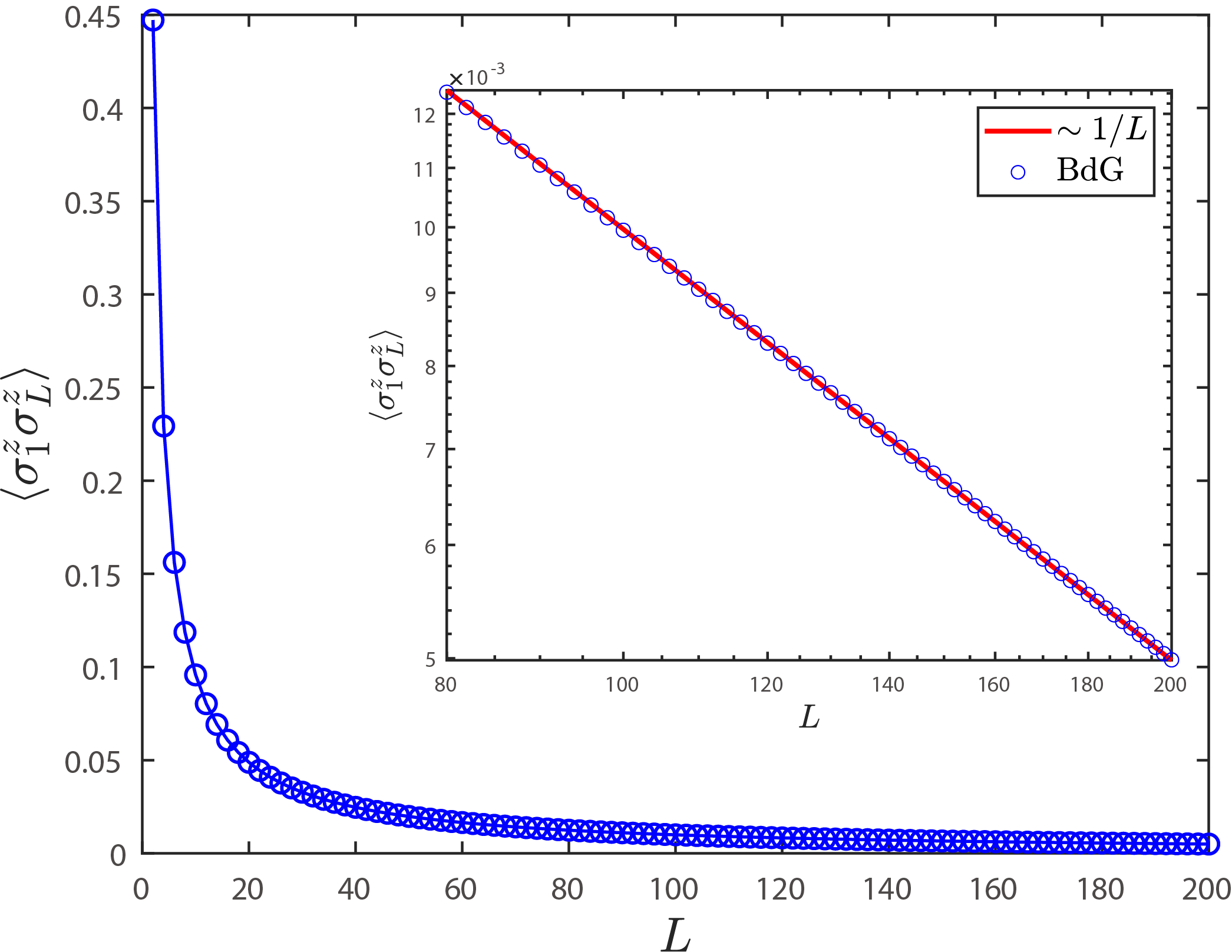}
	\caption{The two point correlation function $\langle\sigma_{1}^{z}\sigma_{L}^{z}\rangle$ is calculated as a function of system size $L$. Inset: A logarithmic plot of the correlation function versus $1/L$ for $L$ going from 80 to 200.}
	\label{fig3}
\end{figure}

\begin{figure}
	\includegraphics[width=8cm]{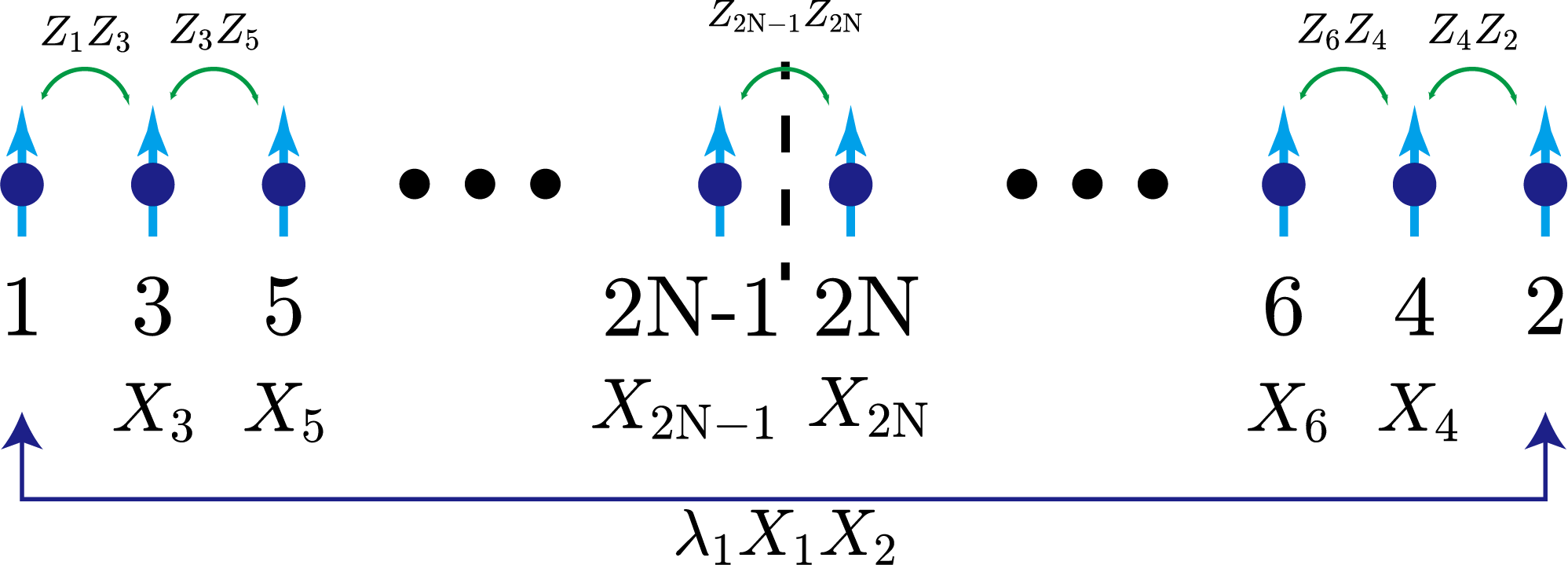}
	\caption{The Hamiltonian after KW transformation on the second chain with $\lambda_2=1$.}
	\label{figrelabel}
\end{figure}
One can relabel the sites as Fig.\ref{figrelabel} and this spin chain becomes a critical lsing chain with a non-local perturbation:
\be
\begin{aligned}
V_{\text{KW}}H_1 V_{\text{KW}}^\dagger= -\sum^{2N-1}_{n=2}\left( X'_{n} + Z'_{n-1} Z'_{n}\right)- Z'_{2N-1} Z'_{2N}-\lambda_1 X'_1 X'_{2}.
\end{aligned}
\ee
We can apply Jordan-Wigner (JW) transformation:
\be
 \begin{aligned}
X'_{n}=i\gamma^1_n \gamma^2_n,\quad Z'_n=\prod^{n-1}_{k=1}(i\gamma^1_k \gamma^2_k)\gamma^1_n
\end{aligned}
\ee
and we can rewrite the spin chain in terms of Majorana fermions :
 \be
 \begin{aligned}
H_{f}=-\sum^{2N-1}_{n=2}\left( i\gamma^1_{n}\gamma^2_{n}+i\gamma^2_{n-1}\gamma^1_{n}\right)-i\gamma^2_{2N-1}\gamma^1_{2N}+\lambda_1 \gamma^1_1\gamma^2_1\gamma^1_{2N}\gamma^2_{2N}.
\end{aligned}
\ee
If $\lambda_1=0$, there are two decoupled Majorana fermions $\gamma^1_1$ and $\gamma^2_{2N}$ and the left Majorana fermion chain has a unique ground state $|\text{GS}.\rangle$. Thus the total fermion chain has two ground states with $i\gamma^1_1\gamma^2_{2N}=\pm 1$. When we turn on $\lambda_1$, the first order perturbation can lift this degeneracy with the gap $\Delta E=2 \lambda_1\langle \text{GS}.|i\gamma^2_1\gamma^1_{2N}|\text{GS}.\rangle\propto \frac{\lambda_1}{L}$.


\end{document}